# Robust topological surface states of Bi$_2$Se$_3$ thin films on amorphous SiO$_2$/Si substrate and a large ambipolar gating effect

Namrata Bansal[1], Nikesh Koirala[2], Matthew Brahlek[2], Myung-Geun Han[3], Yimei Zhu[3], Yue Cao[4], Justin Waugh[4], Daniel S. Dessau[4], and Seongshik Oh[2,a]

[1]Department of Electrical and Computer Engineering, Rutgers, the State University of New Jersey, Piscataway, NJ 08854, USA

[2]Department of Physics and Astronomy, Rutgers Center for Emergent Materials, and Institute for Advanced Materials, Devices and Nanotechnology, Rutgers, the State University of New Jersey, Piscataway, NJ 08854, USA

[3]Condensed Matter Physics & Materials Science, Brookhaven National Laboratory, Upton, NY 11973, USA

[4]Department of Physics, University of Colorado, Boulder, Colorado 80309, USA

a) Corresponding Author, Email: ohsean@physics.rutgers.edu

ABSTRACT:

The recent emergence of topological insulators (TI) has spurred intensive efforts to grow TI thin films on various substrates. However, little is known about how robust the topological surface states (TSS) are against disorders and other detrimental effects originating from the substrates. Here, we report observation of a well-defined TSS on Bi$_2$Se$_3$ films grown on amorphous SiO$_2$ (a-SiO$_2$) substrates and a large gating effect on these films using the underneath doped-Si substrate as the back gate. The films on a-SiO$_2$ were composed of c-axis ordered but random in-plane domains. However, despite the in-plane randomness induced by the amorphous substrate, the transport properties of these films were superior to those of similar films grown on single-crystalline Si(111) substrates, which are structurally better matched but chemically reactive with the films. This work sheds light on the importance of chemical compatibility, compared to lattice matching, for the growth of TI thin films, and also demonstrates that the technologically-important and gatable a-$SiO_2$/Si substrate is a promising platform for TI films.

Topological insulators are a new class of materials that have insulating bulk but metallic surface states that are protected from non-magnetic disorders thanks to their peculiar band structure topology.[1-5] So far, $Bi_2Se_3$ is one of the most extensively studied TI because of its relatively large band gap of 0.3 eV and a well-defined single Dirac cone at the momentum zero in k space.[6] $Bi_2Se_3$ has a layered structure along the c-axis direction of the rhombohedral structure, where unit layers of Bi atoms are sandwiched by Se atoms forming a sequence of Se-Bi-Se-Bi-Se called a quintuple layer (QL). Atoms within a unit layer or QL are strongly bonded, while these QLs are held together by weak van der Waals forces.[7-8] In order to utilize the unique properties of these TSS, numerous efforts have been made to grow $Bi_2Se_3$ films on various substrates.[9-23] Among these, a-$SiO_2$/Si substrate stands out the most due to its unique standing in modern electronics and ease for back-gating. However, due to the complete lack of epitaxial template from the amorphous substrate, there have been only limited efforts to utilize a-$SiO_2$/Si substrates for TI films,[23] and many key questions including whether TSS survives on such an amorphous substrate remain unknown. Here, we show that even if TI $Bi_2Se_3$ films grown on a-

SiO$_2$/Si substrates are plagued by random in-plane domains, they exhibit well-defined TSS in angle resolved photoemission spectroscopy (ARPES) – pushing the limit of topological protection mechanism of TSS. The layered structure of Bi$_2$Se$_3$ allows mitigation of the lattice matching constraint; the weak bonding between the layers allows the epilayer lattice constant to relax to its bulk value with no strain at the interface between the substrate and the film. On these films, we also demonstrate a large (~500%) ambipolar gating effect using the underlying doped Si substrate as a back gate. Surprisingly, these films exhibited superior transport properties to similarly-grown films on single-crystalline Si(111) substrates, which are structurally better matched but chemically more reactive with the films.

The thin Bi$_2$Se$_3$ films were grown on 3-inch oxidized silicon wafers (400 nm SiO$_2$/Si) in a custom-designed SVTA MOS-V-2 MBE system. The base pressure of the system was ~1 × 10$^{-10}$ Torr. Bi and Se fluxes were provided by Knudsen cells and were calibrated using a quartz crystal microbalance and Rutherford backscattering spectroscopy. Surface of the Bi$_2$Se$_3$ film was monitored *in-situ* by reflection high energy electron diffraction (RHEED) and the images of the diffraction patterns were recorded by a digital camera. The oxidized silicon wafers were cleaned *ex-situ* by exposing the substrate to UV-ozone for 5 min before mounting in the MBE chamber to burn off the majority of organic compounds that may be present on the surface. Second cleaning step employed was to heat the substrates up to 300 $^o$C in oxygen pressure of 10$^{-6}$ Torr for 15 min.

The amorphous nature of silicon dioxide is evident in the diffused RHEED pattern (Figure 1a); evolution of the film surface during growth, as monitored by RHEED, is shown in Figure 1. Due to the amorphous template, there is no preferred crystal orientation for Bi$_2$Se$_3$ to grow and thus the first 3 QLs of Bi$_2$Se$_3$ grown at 110 $^o$C resulted in randomly oriented crystallites; RHEED shows a set of concentric ring-like pattern indicating polycrystalline growth (Figure 1b). As the substrate was slowly annealed from low temperature to the second growth temperature of 220 °C in Se ambience, the RHEED showed a transition from polycrystalline ring pattern to weak streaky pattern indicative of single crystalline-like $Bi_2Se_3$ structure (Figure 1c). On another 3 QL deposition at 220 °C, sharp streaky pattern is observed (Figure 1d). The RHEED pattern became sharper with further growth (Figure 1e). However, in contrast to the RHEED patterns of the $Bi_2Se_3$ films grown on Si(111)[9] and $Al_2O_3$(0001)[17] with the same two-step process, which show six-fold symmetry with sample rotation, the RHEED pattern observed here remains identical against sample rotation. This implies that the films on a-$SiO_2$ are composed of random in-plane domains; this is also evidenced in the atomic force microscopy (AFM) image in Figure 1f, which exhibit randomly-oriented triangular islands.

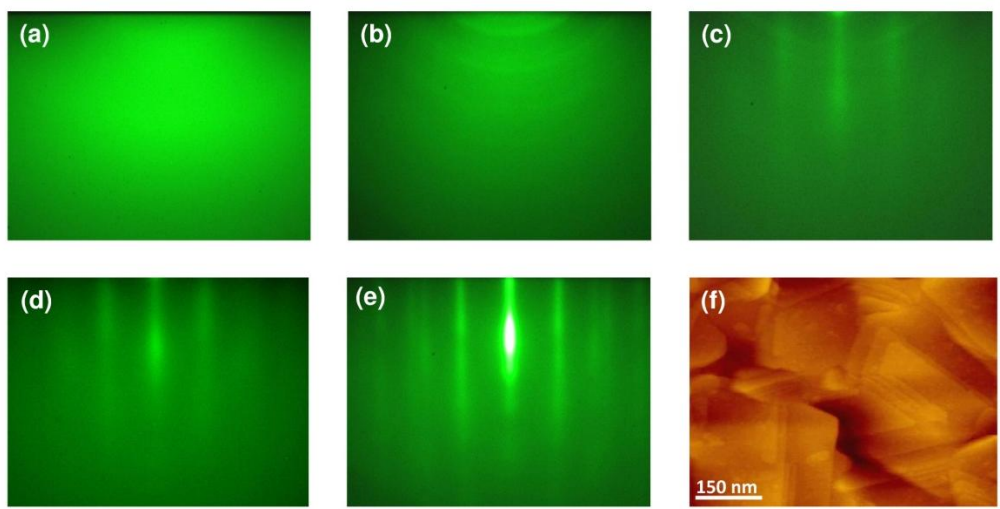

**Figure 1.** Evolution of RHEED pattern during growth of $Bi_2Se_3$ film on a-$SiO_2$ substrates. (a) Amorphous $SiO_2$ substrate. (b) After deposition of 3 QL at 110 °C, a polycrystalline ring-like pattern is seen. (c) The diffraction pattern evolves to streaky pattern as the film is annealed to 220 °C. (d) Streaky pattern indicative of a single-crystalline structure improves on deposition of another 3 QL at 220 °C. (e) Final RHEED pattern of a 50 QL film. The RHEED pattern is rotationally invariant, implying the film is composed of random in-plane domains. (f) AFM image of a 60 QL film grown on a-$SiO_2$. It is composed of randomly orientated triangular islands with QL-step terraces, which is consistent with the RHEED pattern of rotational invariance.

X-ray diffraction (XRD), scanning transmission electron microscopy (STEM) and ARPES studies were carried out to further characterize the films. The XRD measurements were carried out using a Nonius FR571 rotating-anode generator with a copper target and a graphite monochromator, giving a wavelength of 1.5418 Å, and with a Bruker HiStar multi-wire area detector. The STEM sample was prepared by focused ion beam with final Ga$^+$ ion energy of 5 keV. A JEOL ARM 200CF equipped with a cold field-emission gun and double spherical-aberration correctors operated at 200 kV were used for high-angle annular-dark-field (HAADF) STEM. The collection angles for HAADF detectors were 68 to 280 mrad. The ARPES experiments were carried out at the beam-line 12.0.1 of the Advanced Light Source, Lawrence Berkeley National Lab. The film was protected with a Se overlayer after growth and decapped *in-situ* by heating in the final vacuum environment of the analysis chamber.

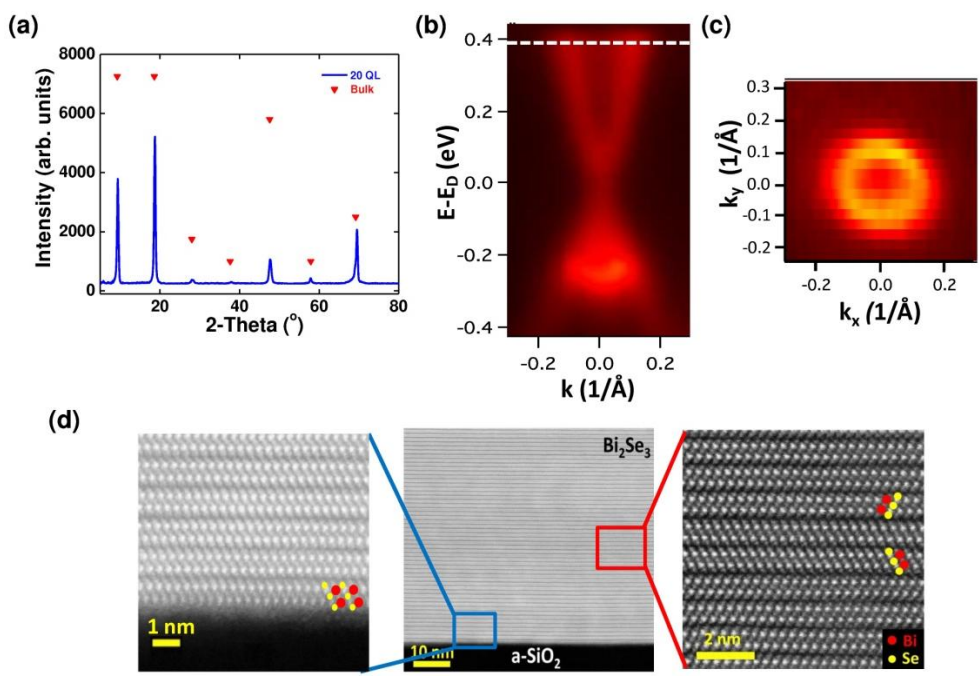

**Figure 2.** XRD, ARPES and STEM images of $Bi_2Se_3$ films on a-$SiO_2$ substrates. (a) XRD scan for 20 QL film. (b) ARPES spectrum for a 60 QL film. (c) Cross-sectional plot of (b) in the $k_x$-$k_y$ plane at the energy level shown as the dotted line in (b). (d) HAADF-STEM cross-section image of the whole structure of a 60 QL film. All the images show nice c-axis layering without any disruption in the QL sequence. The left zoomed-in image shows the interface between the $Bi_2Se_3$ film and the a-$SiO_2$ substrate; even the first QL is well defined other than the very bottom Se atoms. The right zoomed-in image shows well-defined Se-Bi-Se-Bi-Se ordering within each quintuple layer and occasional stacking faults between quintuple layers due to weak interlayer coupling.

In Figure 2a, the XRD θ-2θ scan of a 20 QL $Bi_2Se_3$ film exhibits only (0 0 3n) diffraction peaks, which implies that the film is fully ordered along the c-axis. This is also consistent with the cross-sectional STEM images shown in Figure 2d, which reveal atomically ordered structure down to the first quintuple layer. Individual $Bi_2Se_3$ QLs with full c-axis ordering except for occasional stacking faults between QLs are clearly visible. The interface between $Bi_2Se_3$ and a-$SiO_2$ is sharp with the atomic layer sequence maintained down to the first quintuple layer, as highlighted by spherical dots in Figure 2d, and the amorphous nature of a-$SiO_2$ does not impact the growth along the c-direction beyond the first QL. In Figure 2b, ARPES, clearly shows the V-shaped Dirac surface band; the constant energy Fermi surface map in the $k_x$-$k_y$ plane around Γ-point in Figure 2c shows a ring-like feature formed by the surface states.

The wafer was cut into ~1 cm × 1 cm square substrates and the transport measurements were then carried out in the standard van der Pauw geometry. Temperature dependence of sheet resistance (Figure 3a) exhibits fully metallic behavior for films thicker than 10 QL. However, films thinner than ~10 QL develop an upturn in the sheet resistance at low temperatures due to low mobilities of these ultrathin samples. This implies that even if the films exhibit full c-axis ordering down to the first quintuple layer, the presence of random domains and correspondingly reduced mobilities leaves this system in a weakly insulating phase in this ultrathin regime despite the presence of the topological surface states as confirmed by the ARPES measurement in Figure 2b-c. Figure 3b shows the sheet carrier densities ($n_{2D}$) and respective mobilities extracted from Hall effect data from the slope of $R_{xy}$ vs. B (±0.6 T); $n_{2D}$ changes from ~0.5 × $10^{13}$ $cm^{-2}$ (6 QL) to 3.6 × $10^{13}$ $cm^{-2}$ (100 QL) while the mobility increases fast with film thickness up to 20 QL and remains constant above ~20 QL at ~2000 $cm^2$/Vs.

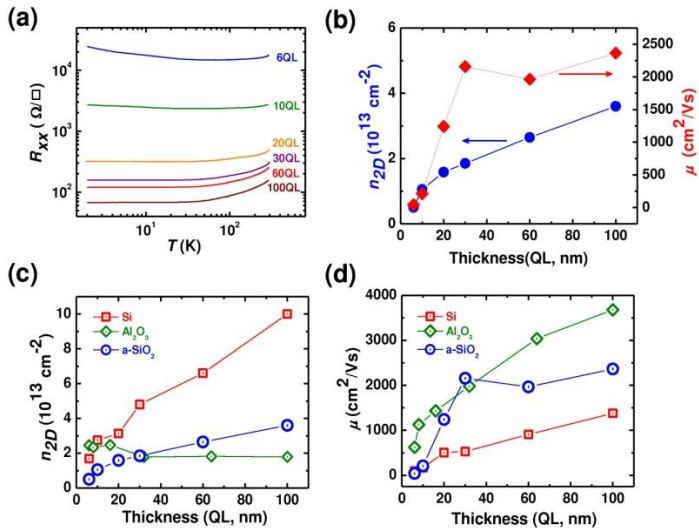

**Figure 3.** Transport properties of $Bi_2Se_3$ films grown on a-$SiO_2$ (a and b); and their comparison with those on Si(111) and $Al_2O_3$(0001) (c and d) (T = 1.5 K, B = ±0.6 T). (a) Resistance vs. temperature for different thicknesses for a-$SiO_2$. (b) Sheet carrier densities (circles) and mobilities (diamonds) vs. thickness for a-$SiO_2$. (c) Sheet carrier densities and (d) mobilities vs. thickness for different substrates. Carrier densities of $Bi_2Se_3$ grown on a-$SiO_2$ are always lower than those on Si(111). Moreover, although the mobilities of $Bi_2Se_3$ grown on a-$SiO_2$ are comparable to those on Si(111) for thickness less than 10 QL, they quickly surpass those on Si substrates at larger thicknesses.

Figure 3c-d compares these films with those grown on single-crystalline substrates, Si(111)[10] and $Al_2O_3$(0001)[17]; all films were grown under nominally identical growth conditions using the same two-temperature growth process[9]: for consistency, all the sheet carrrier densities and mobilities are obtained from the same temperature (1.5 K) and range of magnetic field (±0.6 T). The first thing to notice is that films grown on a-$SiO_2$ have lower sheet carrier densities than those on Si(111) over the entire thickness range (Figure 3c); this implies that both surface and bulk defect densities of the films on a-$SiO_2$ substrates are lower than those on Si(111). On comparison to those on $Al_2O_3$(0001), the films on a-$SiO_2$ exhibit lower carrier densities in the thin regime ($\leq$ 30 QL) but higher values for thicker films. Mobility comparison in Figure 3d shows that $Bi_2Se_3$ films grown on both $Al_2O_3$(0001) and a-$SiO_2$ exhibit superior values to those on Si(111). Only in the ultrathin regime ($\leq$ 10 QL), the mobilities exhibited by the films on Si(111) and a-$SiO_2$ are comparable. This comparison clearly shows that with this standard

growth recipe, the amorphous $SiO_2$ substrate is a superior template than the single-crystalline Si(111) for the transport properties of $Bi_2Se_3$ thin films.

The very observation that a-$SiO_2$ provides better template than Si(111) for the transport properties of $Bi_2Se_3$ is surprising.  From structural point of view, $Bi_2Se_3$ films on a-$SiO_2$, even if they exhibit nice c-axis ordering as shown in Figure 2d, are not as good as those on Si(111) substrates, which are fully single crystalline along both c-axis and ab-plane as shown in Ref. 9.  Therefore, the better transport properties observed on a-$SiO_2$ than on Si(111) cannot be of structural origin.  Although further studies will be needed to clarify this, one plausible scenario is as follows.  Unlike $SiO_2$ and $Al_2O_3$ substrates, Si is chemically reactive and is a dopable semiconductor.  Therefore, even if our two-step growth method suppresses any chemical reaction at the interface, it may still be possible for $Bi_2Se_3$ films and Si substrates to dope each other through an atomistic diffusion process, which would lead to increased carrier densities and reduced mobilities, whereas such a process is forbidden on the inert oxide substrates such as $SiO_2$ and $Al_2O_3$.  Similar process could occur between $Bi_2Se_3$ and any other chemically reactive semiconductor substrates.  For example, transport properties of $Bi_2Se_3$ films grown on SiC(0001) substrate are known to be complicated by the graphene-like conducting channel at the interface.[19]  Similarly, the anomalous transport properties previously observed in $Bi_2Se_3$ films on CdS substrates could be due to such an inter-diffusion process.[21-22]  Another problem with these dopable semiconductor substrates is that once inter-diffusion or some other electronically active process occurs between the film and the substrate, it is hard to distinguish which of the transport contributions are intrinsic to the TI films and which are not.  This observation suggests that for layered TI materials, chemical compatibility between the films and the substrate is a more (or at least, no less) important factor than the lattice matching for their transport properties.

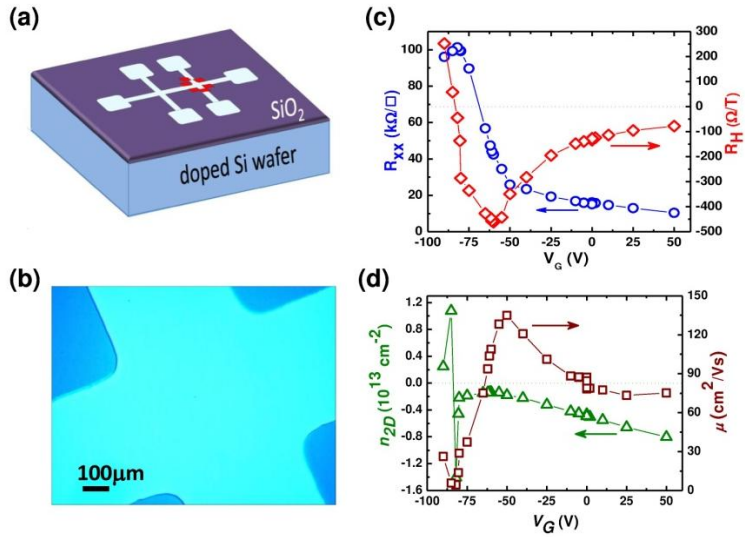

**Figure 4.** Electrostatic back-gating of an 8 QL $Bi_2Se_3$ film on a-$SiO_2$/doped-Si substrate (T = 6 K, B = ±0.6 T). (a) Schematic for the Hall bar device. Optical image of the area enclosed by the red dashed square is shown in (b). (c) Longitudinal resistance and Hall resistance, (d) corresponding sheet carrier densities and mobilities of the back-gated 8 QL film as a function of the gate voltage.

Finally, the $Bi_2Se_3$ films grown on a-$SiO_2$ on doped-Si substrates allowed an effective back-gating. For gating, a 1 cm × 1 cm piece of an 8 QL film was cut out and then ion milled in argon plasma using a shadow mask to make the Hall bar device, and the highly-doped Si substrate was used as the back gate. Figure 4a shows the device schematic with Figure 4b showing an optical image of a part of the device; the Hall bar width is 0.56 mm and the length measured between the mid points of the longitudinal voltage probes is 2.1 mm. Figure 4c shows the sheet resistance, $R_{xx}$, and Hall coefficient, $R_H$, as a function of applied gate voltage, $V_G$; the corresponding sheet carrier density and mobility are shown in Figure 4d. We found that for $Bi_2Se_3$ films grown on $SiO_2$, air exposure tends to reduce both carrier concentration and mobility of the sample. For this sample, in the absence of any gate voltage, the sheet carrier density was $5 \times 10^{12}$ $cm^{-2}$, lower than the fresh samples of similar thicknesses shown in Figure 3 and the sheet resistance ($R_{xx}$) was 15.22 kΩ/sq. about half of the quantum resistance ($h/e^2 \approx$ 25.8 kΩ). When positive gate bias was applied, $R_{xx}$ decreased as more n-type carriers were injected into the film. On the other hand, with negative gate bias, the resistance increased as the n-type carriers were depleted, surpassing the quantum resistance of 25.8 kΩ at $V_G$ = -50 V. When the negative gate bias passed $V_G$ = -82 V, $R_{xx}$ reached a peak of ~100 kΩ/sq and the Hall resistance changed its sign to p-type, indicating that the system passed the Dirac point at this gate voltage.

In summary, we have shown that $Bi_2Se_3$ films grown on amorphous $SiO_2$ substrates, despite their random in-plane disorders, exhibit a well-defined Dirac surface band, superior transport properties to those on single-crystalline Si(111) substrates, and a large ambipolar electric field effect using the doped Si susbtrate as the back gate. This study suggests that chemical compatibility (or inertness) of the substrate should be considered more seriously than lattice matching when choosing substrates for TI materials, and demontrates that the gatable a-$SiO_2$/Si substrate is a promising platform for TI electronics.

ACKNOWLEDGEMENT

This work is supported by National Science Foundation (NSF DMR-0845464) and Office of Naval Research (ONR N000141210456). The electron microscopy work is supported by the U.S. Department of Energy, Office of Basic Science, Division of Materials Science and Engineering, under the Contract number DE-AC02-98CH10886. STEM sample preparation was carried out at the Center for Functional Nanomaterials, Brookhaven National Laboratory. The ARPES work is supported by the National Science Foundation (NSF DMR-1007014), with the Advanced Light Source supported by the Director, Office of Science, Office of Basic Energy Sciences, of the U.S. Department of Energy under Contract No. DE-AC02-05CH11231.

REFERENCES

1      J. E. Moore, *Nature* **464**, 194-198 (2010).

2      M. Z. Hasan, and C. L. Kane, *Rev. Mod. Phys.* **82**, 3045-3067 (2010).

3      X-L. Qi, and S-C. Zhang, *Phys. Today* **63**, 33-38 (2010).

4      P. Roushan, J.Seo, C. V. Parker, Y. S. Hor, D. Hsieh, D. Qian, A. Richardella, M. Z. Hasan, R. J. Cava, and A. Yazdani, *Nature* **460**, 1106-1109 (2009).

5      J. Moore, and L. Balents, *Phys. Rev. B* **75**, 121306(R) (2007).

6      H.-J. Zhang, C.-X. Liu, X.-L. Qi, X.-Y. Deng, X. Dai, S.-C. Zhang, and Z. Fang, *Phys. Rev. B* **80**, 085307 (2009).

7      H. Okamoto, *J. Phase Equilib.* **15**, 195-201 (1994).

8      H. Lind, S. Lidin, and U. Haussermann, *Phys. Rev. B* **72**, 184101 (2005)

9      N. Bansal, Y-S. Kim, E. Edrey, M. Brahlek, Y. Horibe, K. Iida, M. Tanimura, G-H. Li, T. Feng, H-D. Lee, T. Gustafsson, E. Andrei, and S. Oh, *Thin Solid Films* **520**, 224-229 (2011).

10     Y-S. Kim, M. Brahlek, N. Bansal, E. Edrey, G. A. Kapilevich, K. Iida, M. Tanimura, Y. Horibe, S-W. Cheong, and S. Oh, *Phys. Rev. B* **84**, 073109 (2011).

11     G. Zhang, H. Qin, J. Teng, J. Guo, Q. Guo, X. Dai, Z. Fang, and K. Wu, *App. Phys. Lett.* **95**, 053114 (2009).

12     H. D. Li, Z. Y. Wang, X. Kan, X, Guo, H. T. He, Z. Wang, J. N. Wang, T. L. Wong, N. Wang, and M. H. Xie, *New J. Phys.* **12**, 103038 (2010).

13     Z. Y. Wang, H. D. Li, X. Guo, W. K. Ho, and M. H. Xie, *J. Crys. Growth* **334**, 96-102 (2011).

14  J. Chen, H. J. Qin, F. Yang, J. Liu, T. Guan, F. M. Qu, G. H. Zhang, J. R. Shi, X. C. Xie, C. L. Yang, K. H. Wu, Y. Q. Li, and L. Lu, *Phys. Rev. Lett.* **105**, 176602 (2010).

15  A. Richardella, D. M .Zhang, J. S. Lee, A. Koser, D. W. Rench, A. L. Yeats, B. B. Buckley, D. D. Awschalom, and N. Samarth, *App. Phys. Lett.* **97**, 262104 (2010).

16  N. V. Tarakina, S. Schreyeck, T. Borzenko, C. Schumacher, G. Karczewski, K. Brunner, C. Gould, H.Buhmann, and L. W. Molenkamp, *Crys. Gro. & Des.* **12**, 1913-1918 (2012).

17  N. Bansal, Y-S. Kim, M. Brahlek, E, Edrey, and S. Oh, *Phys. Rev. Lett.* **109**, 116804 (2012).

18  A. A. Taskin, S. Sasaki, K. Segawa, and Y. Ando, *Phys. Rev. Lett.* **109**, 066803 (2012).

19  C-Z. Chang, K. He, L-L. Wang, X-C. Ma, M-H. Liu, Z-C. Zhang, X. Chen, Y-Y. Wang, and Q-K. Xue, *Spin* **01**, 21-25 (2011).

20  L. Zhang, R. Hammond, M. Dolev, M. Liu, A. Palevski, and A. Kapitulnik, *App. Phys. Lett.* **101**, 153105 (2012).

21  X. F. Kou, L. He, F. X. Xiu, M. R. Lang, Z. M. Liao, Y. Wang, A. V. Fedorov, X. X. Yu, J. S. Tang, G. Huang, X. W. Jiang, J. F. Zhu, J. Zou, and K. L. Wang, *App. Phys. Lett.* **98**, 242102 (2011).

22  L. He, F. Xiu, X. Yu, M. Teague, W. Jiang, Y. Fan, X. Kou, M. Lang, Y. Wang, G. Huang, N-C. Yeh, and K. L. Wang *Nano Lett.,* **12**, 1486-1490 (2012).

23  S-K. Jeng, K. Joo, Y. Kim, S-M. Yoon, J. H. Lee, M. Kim, J. S. Kim, E. Yoon, S-H. Chun, and Y-S. Kim *Nanoscale*, **5**, 10618-10622 (2013).